\documentclass[12pt]{iopart}
\begin{document}

\title[Electron cyclotron current drive
efficiency]{Electron cyclotron current drive efficiency in an
axisymmetric tokamak}
\author{C. Guti\'errez-Tapia and M. Beltr\'an-Plata}
\address{Departamento de F\'{\i}sica, Instituto Nacional de
Investigaciones Nucleares \\
A. Postal 18-1027, 11801 M\'{e}xico D. F., MEXICO}
\ead{cgt@nuclear.inin.mx}

\begin{abstract}
The neoclassical transport theory is applied to calculate electron
cyclotron current drive (ECCD) efficiency in an axisymmetric
tokamak in the low-collisionality regime. The tokamak ordering is
used to obtain a system of equations that describe the dynamics of
the plasma where the nonlinear ponderomotive (PM) force due to
high-power RF waves is included. The PM force is produced around
an electron cyclotron resonant surface at a specific poloidal
location. The ECCD efficiency is analyzed in the cases of first
and second harmonics (for different impinging angles of the RF
waves) and it is validated using experimental parameter values
from TCV and T-10 tokamaks. The results are in agreement with
those obtained by means of Green's function techniques.
\end{abstract}

\maketitle

\section{Introduction}

Electron cyclotron waves can efficiently drive a localized
non-inductive current in toroidal devices for a number of
applications. The main of these are found in the neoclassical
tearing mode control \cite{1_n}, the fully non-inductive current
drive in tokamaks \cite{2_n, 3_n} and the bootstrap current
compensation in stellerators \cite{4_n}. ECCD results from the
sensitive heating of electrons travelling in one direction in
order to decrease their collision frequency, and thus enhance
their contribution to the toroidal current, compared to their
unheated counterparts moving in the opposite direction \cite{5_n}.
For an off-axis current drive, this current drive mechanism is
offset by the mirror trapping of electrons in toroidal geometries
that drives current in the reverse direction \cite{6_n}. The ECCD
efficiency is usually calculated through a bounce-averaged
quasilinear Fokker-Planck treatment \cite{7_n, 8_n}.

Electron cyclotron (EC) waves have recently attracted a great
interest. Such waves exhibit the very important property of being
able to be excited at a localized particular magnetic surface.
This mechanism considers the introduction of EC waves at a minimum
of the magnetic field, where the resonance condition $\upsilon
_{zr}=\left( \omega -\omega _{Be}\right) /k_{z}$ holds. Many
tokamak experiments have reported that the ECCD efficiency
decreases as the power deposition location is moved away from the
plasma center, either by varying the magnetic field strength
\cite{9_n} or by changing the poloidal steering of the ECCD
launcher \cite{10_n}.

In the low power limit, the ECCD can be calculated from the
relativistic, linearized Fokker-Planck equation using ray tracing
codes \cite{11_n}. If the effects of the radiofrequency (RF)
quasilinear diffusion and the parallel electric field are
included, the bounce-averaged, quasilinear Fokker-Planck codes can
be used \cite{12_n}. However, the nonlinearities associated to
high power effects are not considered. In the present paper we
analyze the ECCD efficiency in an axisymmetric tokamak, in the low
collisionallity regime within the neoclassical transport theory.
The ECCD is calculated including a ponderomotive force
$\mathcal{F}$. The tokamak ordering is used to obtain a system of
equations that describe the dynamics of the plasma where the
nonlinear ponderomotive (PM) force due to high-power RF waves is
included. The PM force is produced around an electron cyclotron
resonant surface at a specific poloidal location. The ECCD
efficiency is analyzed in the cases of the first and second
harmonics (for different impinging angles of the RF waves) and it
is validated using experimental parameter values from TCV and T-10
tokamaks. The results obtained are in agreement with those
delivered by the linearized Fokker-Planck equation.

\section{Basic equations.}

Let us assume a plasma which contains only charged and neutral
particles in a toroidal axysimmetric magnetic field. The
hydrodynamic description of the plasma is taken in the
neoclassical fluid approximation. In this approach, the continuity
equation for the averaging quantities respecting the RF field
becomes
\begin{equation}
\frac{\partial n_\alpha }{\partial t}+\nabla \cdot \left( \eta
_\alpha \mathbf{U}_\alpha \right) =S_{n\alpha }, \label{1}
\end{equation}
where $S_{n\alpha }$ is the source term obeying the condition
\begin{equation}
\sum_{\alpha }q_{\alpha }S_{n\alpha }=\sum_{\alpha }q_{\alpha
}\int \left( \frac{\partial f_\alpha }{\partial t}\right)
_{s0}d\mathbf{v}=0.  \label{2}
\end{equation}
where $S_{n\alpha }=\int \left( \partial f_\alpha /\partial
t\right) _sd\mathbf{v}.$ For the moment equation, we have
\begin{eqnarray}
\fl m_\alpha \frac{\partial \left( n_\alpha \mathbf{U}_\alpha
\right) }{
\partial t} = -\nabla \cdot \widehat{\mathbf{P}}_\alpha -m_\alpha
\nabla \cdot \left( n_\alpha \mathbf{U}_\alpha \mathbf{U}
_\alpha \right) +\mathbf{R}_{c0}^{\left( \alpha \right) }+
\mathbf{R}_{s0}^{\left( \alpha \right) }  \nonumber \\
\lo+ q_\alpha \left[ n_\alpha \mathbf{E}_0+\frac 1cn_\alpha
\mathbf{U}_\alpha \times \mathbf{B}_0\right] + \mathbf{F}_{v\alpha
}-\nabla \cdot \widehat{\pi }_\alpha , \label{3}
\end{eqnarray}

The subindex $\alpha $ refers to the particle species; $m_\alpha$,
$ n_\alpha $, and $U_\alpha $ are the mass, electric charge, the
density of particles and the velocity of the fluid, respectively.
$F_\alpha $ is the ponderomotive force, $\pi _\alpha $ is the
viscosity tensor and, finally, $p_\alpha $ is the pressure defined
by $p_\alpha =n_\alpha T_\alpha ,$ where $T_\alpha $ is the plasma
temperature. On the other hand,
\begin{eqnarray}
R_{c0}^{\left( \alpha \right) } &= m_\alpha \int \left(
\frac{\partial f_\alpha }{\partial t}\right)
_{c0}d\mathbf{\upsilon };  \label{4}
\\
R_{s0}^{\left( \alpha \right) } &= m_\alpha \int \left(
\frac{\partial f_\alpha }{\partial t}\right)
_{s0}d\mathbf{\upsilon },  \label{5}
\end{eqnarray}
is the force of friction between the particles of species $\alpha
$ with neutrals.

The system of equations (\ref{1}) and (\ref{3}) must be completed
with the Maxwell equations for the averaging quantities
\begin{equation}
\eqalign{\nabla \cdot \mathbf{E}&= 4\pi q_\alpha n_{\alpha 0},
\\
\nabla \times \mathbf{E} &= -\frac 1c\frac{\partial \mathbf{B
}}{\partial t},   \\
\nabla \cdot \mathbf{B} &= 0,   \\
\nabla \times \mathbf{B} &= \frac 1c\frac{\partial \mathbf{E}
}{\partial t}+\frac{4\pi }cq_\alpha n_{\alpha 0}\mathbf{U}_\alpha
. \label{6}}
\end{equation}
By using an standard expansion with respect to the ratio of the
gyroradius and the characteristic length, it follows that
\[
\varepsilon =\frac{\rho _{L\alpha }}L,
\]
where $\rho _{L\alpha }$ is the Larmor radius and $L$ is the
characteristic length \cite{13_n}.

\section{Equations of zeroth and firsth orders.}

The system of equations (\ref{1})-(\ref{3}) is reduced, to zero-th
order terms, to
\begin{eqnarray}
\nabla \cdot n_{\alpha 0}\mathbf{U}_{\alpha 0} &= 0,  \label{7} \\
\nabla \cdot \widehat{\mathbf{P}}_{\alpha } &= q_{\alpha }\left[
n_{\alpha 0}\mathbf{E}_{0}+\frac{1}{c}n_{\alpha 0}\mathbf{U}
_{\alpha 0}\times \mathbf{B}_{0}\right] ,  \label{8}
\end{eqnarray}
and the Maxwell equations become
\begin{equation}
\eqalign{\nabla \cdot \mathbf{E}_{0} &= 0,   \\
\nabla \times \mathbf{E}_{0} &= 0,  \\
\nabla \cdot \mathbf{B}_{0} &= 0,  \\
\nabla \times \mathbf{B}_{0} &= \frac{4\pi }{c}\mathbf{J} _{\alpha
0}.  \label{9}}
\end{equation}
Here, we obtain that $ \mathbf{E}_{0}=-\nabla \Phi _{0}$

In this case, the solution of the system (\ref{7})-(\ref{8}) has
the form
\begin{eqnarray}
U_\alpha ^\psi  &= 0,  \label{10} \\
U_\alpha ^\theta  &= \frac c{B_0^2}\left[ \frac{\partial \Phi
_0}{\partial \psi }+\frac 1{q_\alpha n_{\alpha 0}}\frac{\partial
p_\alpha }{\partial \psi }\right] \frac{B_{0\theta }B_0^\theta
}{JB_0^\zeta }+\lambda \left( \psi
\right) B_0^\theta ,  \label{11} \\
U_\alpha ^\zeta  &= -\frac c{B_0^2}\left[ \frac{\partial \Phi
_0}{\partial \psi }+\frac 1{q_\alpha n_{\alpha 0}}\frac{\partial
p_\alpha }{\partial \psi }\right] \frac{B_{0\zeta }B_0^\zeta
}{JB_0^\theta }+\lambda \left( \psi \right) B_0^\zeta . \label{12}
\end{eqnarray}
where we have introduced the toroidal flux coordinates $\left(
\psi ,\theta ,\zeta \right)$. Within this coordinate system, the
contravariant forms of the magnetic field and of the fluid
velocity for an axisymmetric tokamak are written as
\begin{equation}
\eqalign{\mathbf{B}_0 &= I\left( \psi \right) \nabla \zeta +\nabla
\zeta \times \nabla \psi,
 \\
\mathbf{U}_\alpha  &= U_\alpha ^\theta \mathbf{e}_\theta +U_\alpha
^\zeta \mathbf{e}_\zeta ,  \label{13}}
\end{equation}
where $n_{\alpha 0}=cte$ and the function $\lambda \left( \psi
\right)$ is unknown.

Considering the inequality $B_{0\zeta }B_{0}^{\zeta }\gg
B_{0\theta }B_{0}^{\theta }$ and $B_{0}^{2}\simeq B_{0\zeta
}B_{0}^{\zeta }$ we reduce the equations (\ref{11}) and (\ref{12})
to the form
\begin{eqnarray}
U_{\alpha }^{\theta } &=&\lambda \left( \psi \right) B_{0}^{\theta };
\label{14} \\
U_{\alpha }^{\zeta } &=&-\frac{c}{JB_{0}^{\theta }}\left[
\frac{\partial \Phi _{0}}{\partial \psi }+\frac{1}{q_{\alpha
}n_{\alpha 0}}\frac{\partial p_{\alpha }}{\partial \psi }\right]
+\lambda \left( \psi \right) B_{0}^{\zeta }.  \label{15}
\end{eqnarray}
which are the zero-th order velocity equations. The corresponding
toroidal current density is calculated from the relationship
\begin{eqnarray}
j^{\zeta } &\equiv q_{\alpha }n_{\alpha }U_{\alpha }^{\zeta }
\nonumber \\
&= -\frac{cq_{\alpha }n_{\alpha }}{JB_{0}^{\theta }}\left[
\frac{\partial \Phi _{0}}{\partial \psi }+\frac{1}{q_{\alpha
}n_{\alpha 0}}\frac{\partial p_{\alpha }}{\partial \psi }\right]
+q_{\alpha }n_{\alpha }\lambda \left( \psi \right) B_{0}^{\zeta }.
\label{16}
\end{eqnarray}

Equations (\ref{1})-(\ref{3}) containing terms to first order.
Ignoring the source term they become
\begin{eqnarray}
\frac{\partial n_{\alpha 0}}{\partial t}+\nabla \cdot \left(
n_{\alpha 0} \mathbf{U}_\alpha \right) &= 0,  \label{17} \\
m_\alpha \frac{\partial \left( n_{\alpha 0}\mathbf{U}_\alpha
\right) }{\partial t} &= -\nabla p_\alpha -\nabla \cdot
\widehat{\pi }_\alpha -m_\alpha \nabla \cdot \left( n_{\alpha
0}\mathbf{U}_\alpha \mathbf{U}_\alpha
\right) +\mathbf{R}_\alpha ^{\left( \alpha .n\right) }\nonumber \\
&+ q_\alpha \left[ n_{\alpha 0}\mathbf{E}_0+\frac 1cn_{\alpha 0}
\mathbf{U}_\alpha \times \mathbf{B}_0\right] +
\mathbf{\mathcal{F}}_\alpha ,  \label{18}
\end{eqnarray}
where $\widehat{\mathbf{P}}_\alpha =p_\alpha +\widehat{\pi
}_\alpha , $ $p_\alpha $ is the scalar pressure, $\widehat{\pi
}_\alpha $ is the viscosity tensor, $\mathbf{R}_\alpha ^{\left( \alpha .n\right) }=
\mathbf{R}_{c0}^{\left( \alpha \right) }+\mathbf{R}
_{s0}^{\left( \alpha \right) }$ is the friction force from the
particles of species $\alpha $ with neutrals and $\mathbf{\mathcal{F
}}_\alpha =\mathbf{F}_{v\alpha }-\nabla \cdot \widehat{
\mathbf{\pi }}_\alpha $ is the average ponderomotive force
associated with the RF field acting on the particles.

\section{Steady state equations.}

In a steady state, the above system of equations can be written in
the form
\begin{eqnarray}
\nabla \cdot \left( \eta _{\alpha 0}\mathbf{U}_\alpha \right) & =
& 0, \label{19} \\
m_\alpha \nabla \cdot \left( \eta _{\alpha 0}\mathbf{U}_\alpha
\mathbf{U}_\alpha \right) & =& -\nabla p_\alpha -\nabla \cdot
\widehat{ \pi }_\alpha +\mathbf{R}_\alpha ^{\left( \alpha
..n\right) }
\nonumber \\
&&+q_\alpha \left[ n_{\alpha 0}\mathbf{E}_0+\frac 1cn_{\alpha 0}
\mathbf{U}_\alpha \times \mathbf{B}_0\right] +
\mathbf{\mathcal{F}}_\alpha , \label{20}
\end{eqnarray}
where the $\zeta$ component of the velocity becomes
\begin{eqnarray}
\fl U_\alpha ^\zeta = -\frac c{B_0^2}\left( \frac{\partial \Phi
_0}{\partial \psi }+\frac 1{q_\alpha n_{\alpha 0}}\frac{\partial
p_\alpha }{\partial \psi }\right) \frac{B_{0\zeta }B_0^\zeta
}{JB_0^\theta }  + \frac{B_0^\zeta }{\mathbf{\mu }_0\cdot
\mathbf{B}_0} \left\{ \left\langle \mathbf{B}_0\cdot
\mathbf{\mathcal{F}} _\alpha \right\rangle -k \right. \nonumber
\\
\lo- \left.\frac c{B_0^2}\left( \frac{\partial \Phi _0}{\partial
\psi }+\frac 1{q_\alpha n_{\alpha 0}}\frac{\partial p_\alpha
}{\partial \psi }\right) \left[ \frac{\mu _\theta B_{0\theta
}B_0^\theta }{JB_0^\zeta }-\frac{\mu _\zeta B_{0\zeta }B_0^\zeta
}{JB_0^\theta }\right] \right\} ,  \label{21}
\end{eqnarray}
where $\mu_\theta$ and $\mu_\zeta$ are the poloidal and toroidal
coefficients of viscosity, respectively. In the limit $B_{0\theta
}B_0^\theta <<B_{0\zeta }B_0^\zeta$ and $B_0^2\approx B_{0\zeta
}B_0^\zeta$, the toroidal current density is reduced to
\begin{eqnarray}
\fl j^{(\zeta )} =-\frac{cn_\alpha q_\alpha }{JB_0^\theta }\left[
\frac{\partial \Phi _0}{
\partial \psi }+\frac 1{q_\alpha n_{\alpha 0}}\frac{\partial p_\alpha }{
\partial \psi }\right]  +\frac{n_\alpha q_\alpha B_0^\zeta }
{\left( \mathbf{\mu }_0\cdot
\mathbf{B}_0\right) +m_\alpha n_{\alpha 0}v_{\alpha n}\left\langle
B_0^2\right\rangle }\left\{ \left\langle \mathbf{B}_0\cdot
\mathbf{\mathcal{F}}_\alpha \right\rangle -k\right.  \nonumber \\
\lo+ \left. \frac{c\mu _\zeta }{JB_0^\theta }\left[ \frac{\partial
\Phi _0}{
\partial \psi }+\frac 1{q_\alpha n_{\alpha 0}}\frac{\partial p_\alpha }{
\partial \psi }\right]
 +m_\alpha n_{\alpha 0}v_{\alpha n}\frac{c\left\langle B_{0\zeta
}\right\rangle }{JB_0^\theta }\left[ \frac{\partial \Phi
_0}{\partial \psi } +\frac 1{q_\alpha n_{\alpha 0}}\frac{\partial
p_\alpha }{\partial \psi } \right] \right\},  \label{22}
\end{eqnarray}
and
\begin{eqnarray}
\fl \lambda \left( \psi \right) =\frac 1{\left( \mathbf{\mu
}_0\cdot \mathbf{B}_0\right) +m_\alpha \eta _{\alpha 0}v_{\alpha
n}\left\langle B_0^2\right\rangle }\left\{ \left\langle
\mathbf{B}_0\cdot \mathbf{\mathcal{F}} _\alpha \right\rangle -k
+\frac{c\mu _\zeta }{JB_0^\theta }\left[ \frac{\partial \Phi _0}{
\partial \psi }+\frac 1{q_\alpha \eta _{\alpha 0}}\frac{\partial p_\alpha }{
\partial \psi }\right] \right.  \nonumber \\
\lo+ \left. m_\alpha \eta _{\alpha 0}v_{\alpha
n}\frac{c\left\langle B_{0\zeta }\right\rangle }{JB_0^\theta
}\left[ \frac{\partial \Phi _0}{\partial \psi } +\frac 1{q_\alpha
\eta _{\alpha 0}}\frac{\partial p_\alpha }{\partial \psi } \right]
\right\} .  \label{23}
\end{eqnarray}

\section{The ponderomotive force.}

The RF ponderomotive force has several representations according
to its functionality with respect to time. In this work, we chose
the following expression of the time averaged ponderomotive force
\cite{14_n}
\begin{equation}
\mathbf{\mathcal{F}}_\alpha =\frac 12Re\left\{ \frac i\omega
\nabla E^{*}\cdot \mathbf{j}_\alpha -\nabla \cdot \left[
\mathbf{j} _\alpha \left( \frac i\omega E^{*}+\frac{4\pi
\mathbf{j}_\alpha ^{*} }{\omega _{p\alpha }^2}\right) \right]
\right\}, \label{24}
\end{equation}
where $\omega$ is the frequency of the RF wave, $\mathbf{j}_\alpha $
is the current density of particles of species $\alpha$ induced by
the RF field $\mathbf{E}$ and $\omega _{p\alpha }^2=4\pi n_\alpha
q_\alpha ^2/m_\alpha $ is the plasma frequency of particles of
species $\alpha$.

Given the Ohm's law, we assume the conductivity tensor $\sigma
_{jk}$, which depends on the assumed characteristic frequency. The
ponderomotive force is introduced thanks to a system of orthogonal
coordinates $\left(\mathbf{e}_1,\mathbf{e}_2,\mathbf{e}_3\right)$
with the components of the conductivity tensor in the form
\begin{equation}
\sigma _{11}=\sigma _{22}=\frac{i\omega }{4\pi }\frac \upsilon
{1-u};\qquad \sigma _{12}=-\sigma _{21}=-\frac \omega {4\pi
}\frac{\sqrt{u}\upsilon }{1-u}, \label{25}
\end{equation}
where $\upsilon =\omega _{pe}^2/\omega ^2,u=\Omega^2/\omega ^2$\
and $\Omega =eB/m_ec$. Here, we have considered the case of an
extraordinary wave ($E_1,E_2,0$).

\begin{table}
\caption{\label{t1}TCV tokamak data for the first and second
harmonics.}

\begin{indented}
\lineup

\item[]\begin{tabular}{@{}llllllll}
\br
&$B_0$ & $n_e$ & $a$ & $R_0$ &$P$ & $T_e$ & $q$ \cr
&(gauss)
&(cm$^{-3}$) & (cm) & (cm) & (MW) & (KeV) &  \cr
\mr
1th harmonic
&$1.43\times 10^4$ & $1.75\times 10^{13}$ & $25.0$ & $88.0$ &
$1.0$ & $3.5$ & $10$ \cr
2nd harmonic &$1.43\times 10^4$ &
$1.75\times 10^{13}$ & $25.0$ & $88.0$ & $1.0$ & 3.5 & $10$ \cr
\br
\end{tabular}
\end{indented}
\end{table}

Thus, the corresponding components of the ponderomotive force take
the form
\begin{eqnarray}
{\mathcal F}_{\alpha 1} &= \frac 1{8 \pi }\frac{\upsilon
Imk_1}{\left( 1-u\right) }\left( 1+u\right) \left| E\right| ^2,
\label{26} \\
{\mathcal F}_{\alpha 2} &= \frac 1{8\pi }\frac{\upsilon
Imk_2}{\left( 1-u\right) ^2}\left( 1+u\right) \left| E\right| ^2.
\label{27}
\end{eqnarray}

Now, by calculating the average value $\left\langle
\mathbf{B}_0\cdot \mathbf{\mathcal{F}}_\alpha \right\rangle $
assuming that
\begin{equation}
\eqalign{\mathbf{B}_0 &= B_0^\theta \widehat{e}_\theta +B_0^\zeta
\widehat{e}_\zeta , \\
\mathbf{\mathcal{F}}_\alpha  &= {\mathcal F}_{\alpha \psi
}\widehat{e} _\psi {\mathcal +F}_{\alpha
\theta}\widehat{e}_\theta,  \label{28}}
\end{equation}
and the equation,
\begin{equation}
\mathbf{B}_0\cdot \mathbf{\mathcal{F}}_\alpha ={\mathcal F}
_{\alpha \theta }B_0^\theta ,  \label{29}
\end{equation}
and finally substituting (\ref{26})-(\ref{28}) in (\ref{29}) and
averaging, we obtain
\[
\left\langle \mathbf{B}_0\cdot \mathbf{{\mathcal F}}_\alpha
\right\rangle =\frac{B_0Imk_2}{16\pi ^2qR_0}\left[ \frac \upsilon
{\left( 1-u\right) ^2}\left( \left( 1+u\right) \left| E\right|
^2+4\sqrt{u}Im\left( E_1E_2^{*}\right) \right) \right] ,
\]
where it has been considered that
\[
\partial _2\frac{u\upsilon }{\left( 1-u\right) ^2}=\partial
_2\frac{\sqrt{u}\upsilon \left( 1+u\right) }{\left( 1-u\right)
^2}=0,
\]
and the Hamada coordinates \cite{19_n} have been used.

\begin{table}
\caption{\label{t2}T-10 tokamak data for the first and second
harmonics.}

\begin{indented}
\lineup

\item[]\begin{tabular}{@{}llllllll}
\br
&$B_0$ & $n_e$ & $a$ & $R_0$ & $P$ & $T_e$ & $q$ \cr
&(gauss) &
(cm$^{-3}$) & (cm) & (cm) & (MW) & (KeV) & \cr
\mr
1th harmonic &
$2.78\times 10^4$ & $0.54\times 10^{13}$ & $38.7$ & $1.5\times
10^2$ & $0.75$ & $6.3$ & $9$ \cr
2nd harmonic & $2.47\times 10^4$
& $0.54\times 10^{13}$ & $38.7$ & $1.5\times 10^2$ & $0.45$ &
$3.8$ & $9$ \cr
\br
\end{tabular}
\end{indented}
\end{table}

Finally, neglecting the attenuation of the RF wave, we obtain
\begin{equation}
\left\langle \mathbf{B}_0\cdot \mathbf{{\mathcal F}}_\alpha
\right\rangle =\frac{B_0Imk_2}{16\pi ^2qR_0}\frac{\upsilon \left(
1+u\right) }{\left( 1-u\right) ^2}\left| E\right| ^2, \label{30}
\end{equation}
while the current density related to the ponderomotive force, from
(\ref{22}), assuming the steady state,  becomes
\begin{equation}
j_\alpha ^{\zeta (p)} \equiv \left\langle J \right\rangle \approx
\frac{n_\alpha q_\alpha k_2^{''}|E|^2}{32 \pi^3 q R_0^2 m_\alpha
n_{\alpha 0} \nu_{\alpha n}} \frac{v(1+u)}{(1-u)^2} \left( 1 +
\frac{3}{2} \varepsilon \right). \label{31}
\end{equation}

\section{Analysis of results.}

In order to examine the expression for the current density
(\ref{31}) associated with the ponderomotive force, we adopt the
criteria that the deposited energy by the RF wave has to be bigger
that the internal enegy ($E^2>NT$) so to include the nonlinear
effects. This condition is satisfied on Tokamaks TCV ad T-10,
where an analysis reported in \cite{15_n, 16_n} shows that
$E^2\sim 5.75689\times 10^{13}>NT=98122.5$ and $E^2\sim
3.37737\times 10^{13}>NT=54500.5$ in these tokamaks, respectively.
Such results indicate that the effect of the nonlinear
ponderomotive force is highly important in determining the energy
density introduced by the RF wave. We will consider a flux
cylinder with a radius equal to the Larmor radius so to calculate
this energy.

The corresponding data for the TCV and T-10 Tokamaks are
summarized in Table~\ref{t1} and Table~\ref{2} \cite{15_n, 16_n},
respectively. In both reports, the first and second harmonics of
the $EC$ waves were used provided that the introduction of the RF
wave took place at the high magnetic field (HF) side.

The power associated to the amplitude of the electric field
follows from
\begin{equation}
\left| E\right| ^2=\frac{2P}{\omega \varepsilon _0},  \label{32}
\end{equation}
where $P$ is the power of the wave per time unit and $\varepsilon
_0$ is the permittivity in vacuum, assumed to be a constant.

For the imaginary part of the permittivity, $\Im k_2=k_2^{''}$, we
use the expression reported in \cite{17_n}. In the case of the
first harmonic, one has that
\begin{equation}
\frac{k_2^{''}}{k_2}=\frac{2\sqrt{\pi }}{75}\beta _T^2\left(
2-q\right) z^{3/2}e^{-z}\left[ \frac{qz^2}{14}+\frac{\left(
5/2-q\right) ^2}{q\left| F\right| ^2}\right] ,  \label{33}
\end{equation}
where
\begin{eqnarray}
F\left( z\right) &=&\frac 34\left[ \frac 12+z+\sqrt{\pi
}z\sqrt{-z} e^{-z}\left( \Phi \left( \sqrt{-z}\right) -1\right)
\right] , \label{34}
\end{eqnarray}
Here, $\Phi \left( x\right) $ is the error function, $\beta _T=
\upsilon _{T_e}/c$ is the ratio of the thermal velocity to the
light velocity in vacuum, and $z=2\left( \Omega-\omega \right)
/\Omega\beta _T^2$.

Analogously, for the second harmonic, we have
\begin{equation}
\frac{k_{2}^{''}}{k_{2}} = \frac{2^{2}}{2(5!)}\sqrt{\pi }q\left(
1+\Gamma _{1}\right) ^{2}z^{3/2}e^{-z},\label{35}
\end{equation}
where
\[
(1+\Gamma _{1}) = \frac{2^{2}-1-q\left( 1-1/2\right) }{2^{2}-1-q}.
\]

The viscosity is neglected while the collision frequency between
electrons and neutrals was taken in the form
\begin{equation}
\nu _{en}=\tau _e^{-1}=\left( 3.5\times 10^4T^{3/2}/n_e\right)
^{-1}. \label{36}
\end{equation}

It is important to notice that the current density is highly
unstable and it depends strongly on the  $\Omega^2 / \omega ^2$
relationship. However, it is possible to find an interval where
the current density stabilizes and, furthermore, its values
reproduce those experimental ones reported in \cite{15_n, 16_n},
as can be observed in \fref{F1} and \fref{F3}.

From \fref{F2} and \fref{F4}, we observe a process in which the
Fish and Ohkawa mechanisms weaken each other, as reported in
\cite{7_n}. The general behavior is in good agreement with that
described in \cite{8_n, 18_n}, considering that their calculation
was obtained from the linearized Fokker-Planck equation.

Here, the density profile has been modelled as a parabolic one in
order to analyze the current, $ n_e=n_{0e}\left( 1-\frac 23\left(
\frac ra\right) \right) ^2, $ where $a$ is the Tokamak minor
radius.

It can be noticed in \fref{F5} that, to first order terms in the
parameter $\epsilon$, the driven current density increases with
the radius.

\section{Conclusions.}

The development of a driven current density expression that takes
into account the ponderomotive force created by EC waves, has
required the use of the neoclassical transport equations up to
first order terms with respect to the parameter $\varepsilon =\rho
/ L$ at a steady state.

The driven current density has been initially obtained in a system
of toroidal flux coordinates. That description of the current
density is transformed in terms of the Hamada coordinates
\cite{19_n} which is necessary for its validation with
experimental results. Thus, the expression for the ponderomotive
force reported in \cite{14_n}, is written in a local system of
orthogonal coordinates $\left( \mathbf{e}_1,\mathbf{e}_2,
\mathbf{e}_3\right)$, where $\mathbf{e}_3$ is parallel to the
toroidal magnetic field.

The driven current density generated by an extraordinary wave at
the cyclotron resonance of electrons, is analyzed as a function of
the $\Omega^2/\omega ^2$ ratio. This is accomplished by using the
parameters of the TCV and T-10 Tokamaks at the first and second
harmonics, assuming that the introduction of the wave takes place
at the HF side. From this results, we have obtained an interval of
frequencies, in agreement with the experiments, where the current
shows a stable behavior.

 In the particular case of a parabolic profile, it has been
shown that the ECCD increases with the radius, at first
approximation. Finally, it is important to notice that, according
to \cite{17_n}, the efficiency is higher for the second harmonic
as it is shown in \fref{F5}.

\ack

This work is partially supported by Conacyt, Mexico, under
contract 33873-E.

\Figures

\begin{figure}
\caption{\label{F1}Current efficiency plotted against
$\Omega^2/\omega^2$ for the first harmonic with parametric values
from the tokamaks a) T-10 with $f = 140/1.06$ Ghz (solid) and $f =
140/1.12$ Ghz (dashed), and b) TCV with $f = 82.7/1.484$ Ghz
(solid) and $f = 82.7/1.5$ Ghz (dashed)}
\end{figure}

\begin{figure}
\caption{\label{F2}Current efficiency plotted against
$\varepsilon$ for the first harmonic with values taken by
parameters of the tokamaks a) T-10 with $r=0.0$ cm (solid) and
$r=88$ cm (dashed), and b) TCV with $r=0.0$ cm (solid) and $r=88$
cm (dashed).}
\end{figure}

\begin{figure}
\caption{\label{F3}Current efficiency plotted against
$\Omega^2/\omega^2$ for the second harmonic with parameters from
the tokamaks  a) T-10 with $f = 140 \times 0.7$ Ghz (solid) and $f
= 140 \times 0.715$ Ghz (dashed) and b) TCV with $f = 82.7 \times
1.331$ Ghz (solid) and $f = 82.7 \times 1.339$ Ghz (dashed).}
\end{figure}

\begin{figure}
\caption{\label{F4}Current efficiency plotted against
$\varepsilon$ for the second harmonic with parameter values from
the tokamaks a) T-10 with $r=0.0$ cm (solid) and $r=88$ cm
(dashed), and b) TCV with $r=0.0$ cm (solid) and $r=88$ cm
(dashed).}
\end{figure}

\begin{figure}
\caption{\label{F5}Current efficiency plotted against the minor
radius $r$ for the second harmonic with parameters from the
tokamaks a) T-10 and b) TCV.}
\end{figure}


\section*{References}
\begin{thebibliography}{99}
\bibitem{1_n} Isayama A \etal 2000 {\it Plasma Phys. Control. Fusion}
\textbf{42} L37
\bibitem{2_n} Lin-Liu Y R, Chan V S, \etal 2003 {\it Phys. Plasmas}
\textbf{10} 4064
\bibitem{3_n} Prater R 2004 {\it Phys. Plasmas} \textbf{11} 2349
\bibitem{4_n} Castej\'{o}n F \etal 2004 {\it Nucl. Fusion} \textbf{44} 593
\bibitem{5_n} Fish N J and Boozer A H 1980 {\it Phys. Rev. Lett.} \textbf{45}
720
\bibitem{6_n} Ohkawa T 1970 {\it Nucl. Fusion} \textbf{15} 185
\bibitem{7_n} Cordey J G, Edington T \etal 1982 {\it Plasma Phys.}
\textbf{24} 73
\bibitem{8_n} Taguchi M 1989 {\it Plasma Phys. Contr. Fusion} \textbf{31} 241
\bibitem{9_n} Alikaev V V \etal 1995 {\it Nucl. Fusion} \textbf{35} 369
\bibitem{10_n} Sauter O \etal 2001 {\it Phys. Plasmas} \textbf{8} 2199
\bibitem{11_n} Matsuda K 1989 {\it IEEE Trans. Plasma Sci.} \textbf{17} 6
\bibitem{12_n} Harvey R W and McCoy M G  1993 {\it Proc. IAEA Technical
Comittee Meeting (Montreal, 1992)} p~498
\bibitem{13_n} Martinell J J and Guti\'{e}rrez-Tapia C 2001 {\it Phys. Plasmas}
\textbf{8} 2808
\bibitem{14_n} Klima R 1998 {\it Czech. J. Phys.} \textbf{B 18} 1280
\bibitem{15_n} Alikaev V V \etal 1992 {\it Nucl. Fusion} \textbf{32} 1811
\bibitem{16_n} Sauter O \etal 2000 {\it Phys. Rev. Lett.} \textbf{84} 3322
\bibitem{17_n} Litvak A G Ed. 1992 {\it High-frequency Plasma Heating} (New York: AIP
press) p~1
\bibitem{18_n} Cohen R H 1987 {\it Phys. Fluids} \textbf{30} 2442
\bibitem{19_n} Coronado M and Talmadge J N 1993 {\it Phys. Fluids} \textbf{B
5} 1200

\end{thebibliography}
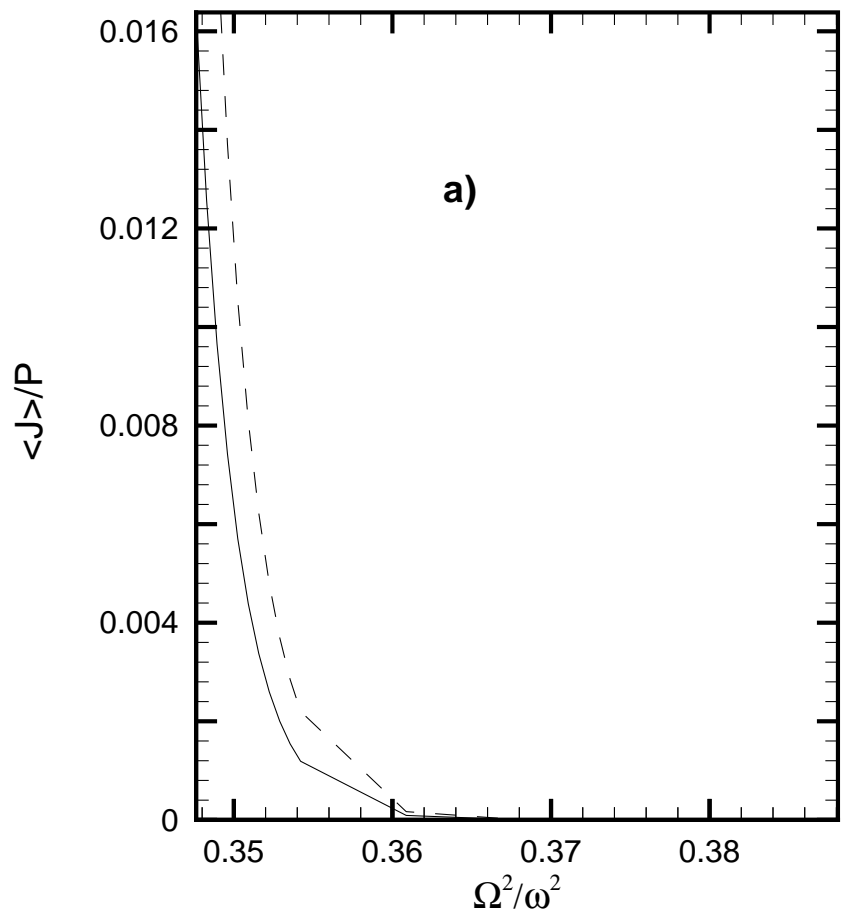
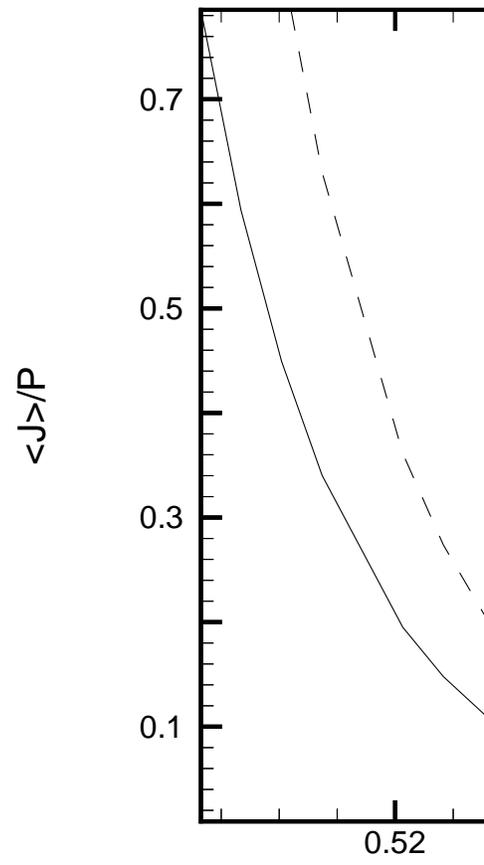
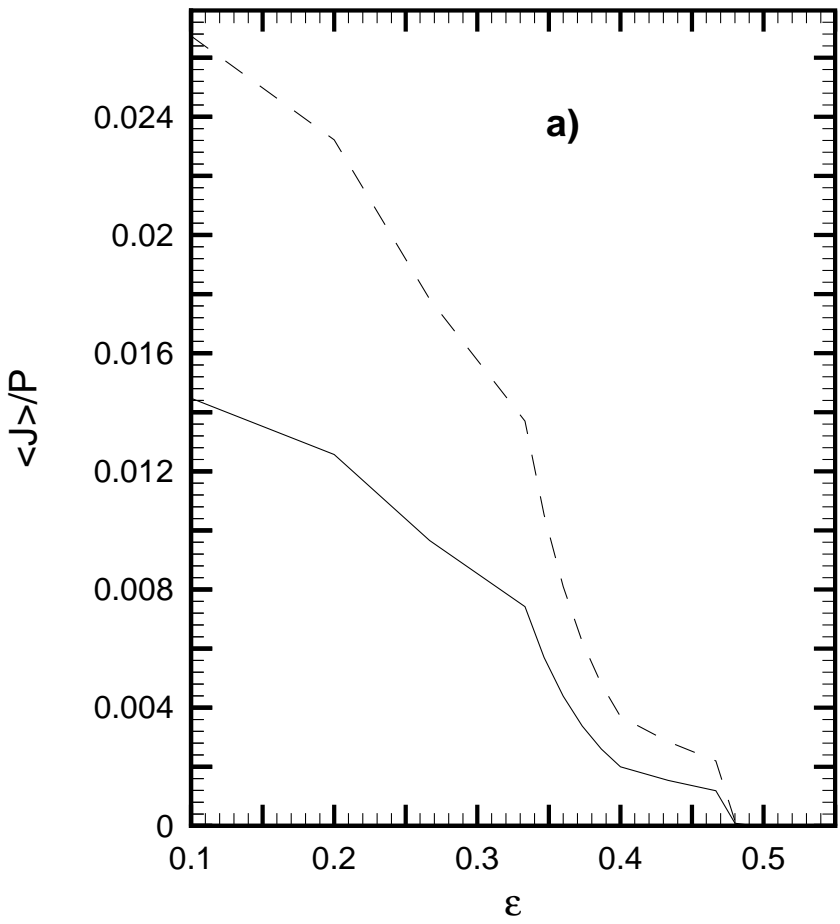
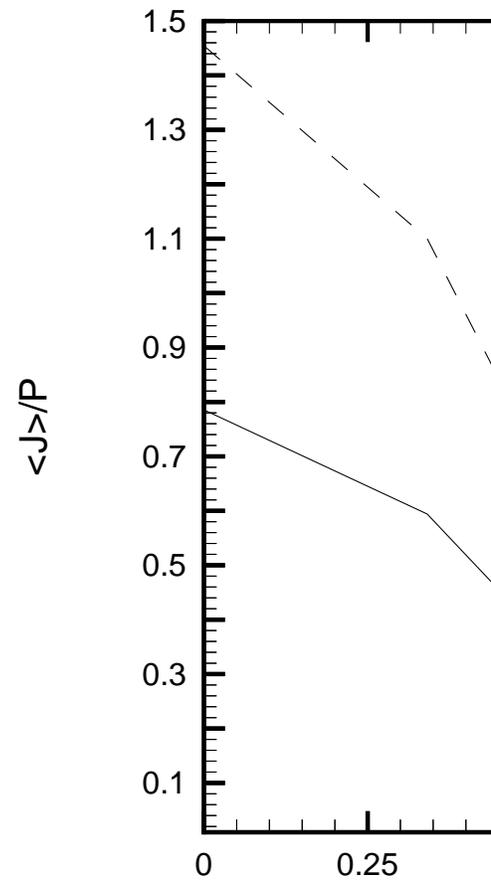
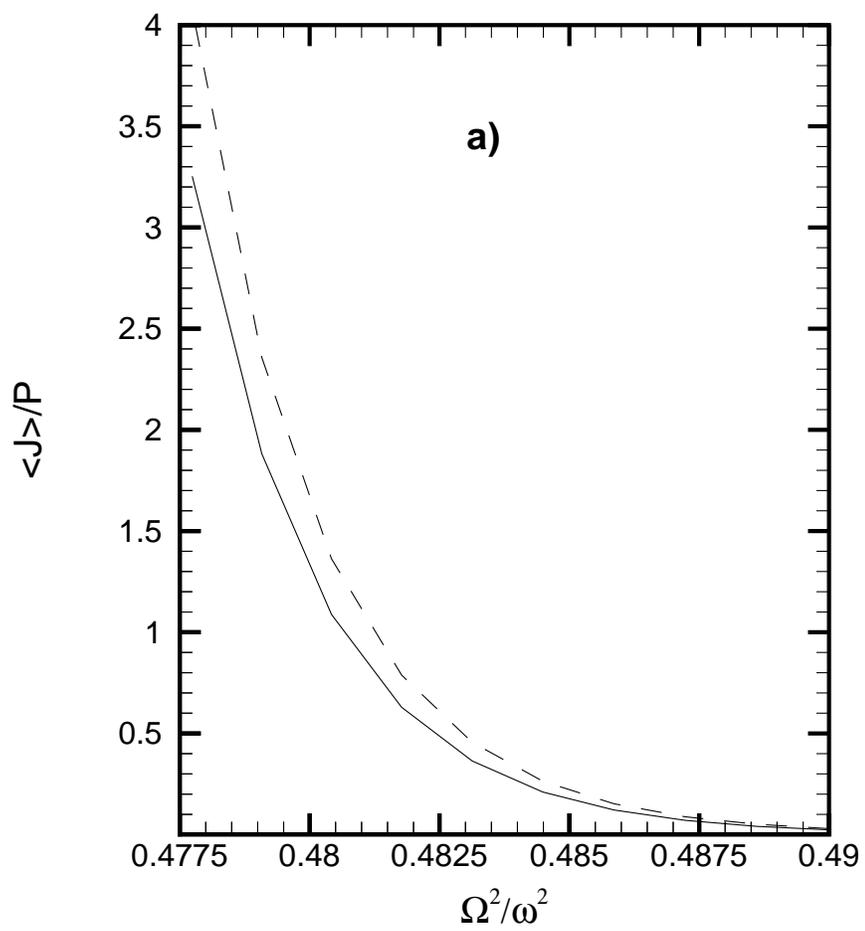
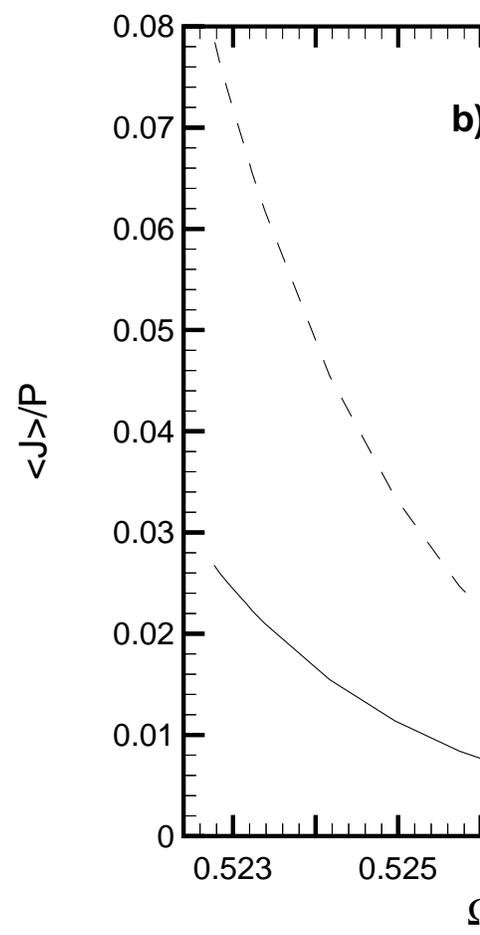
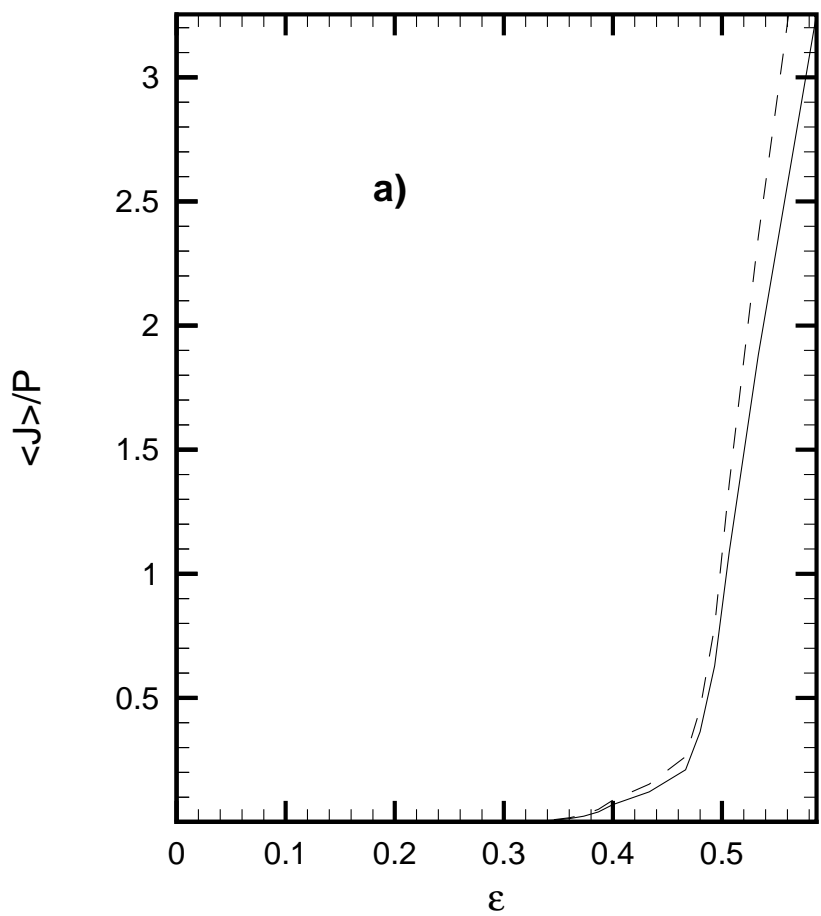
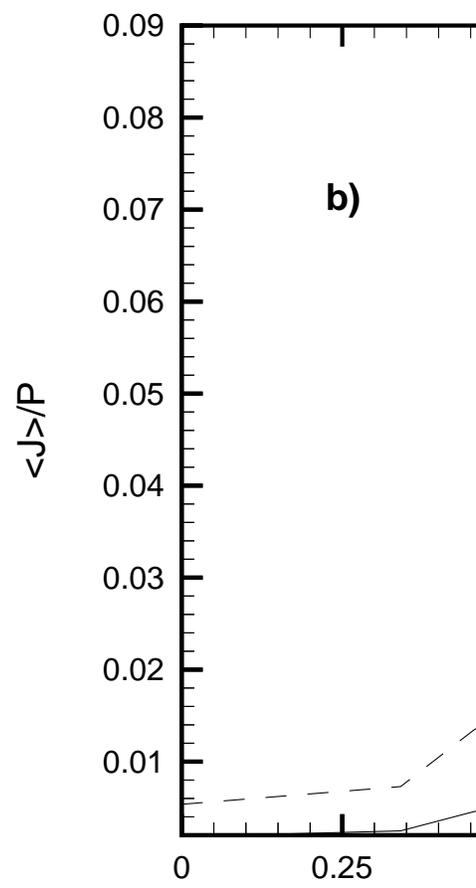
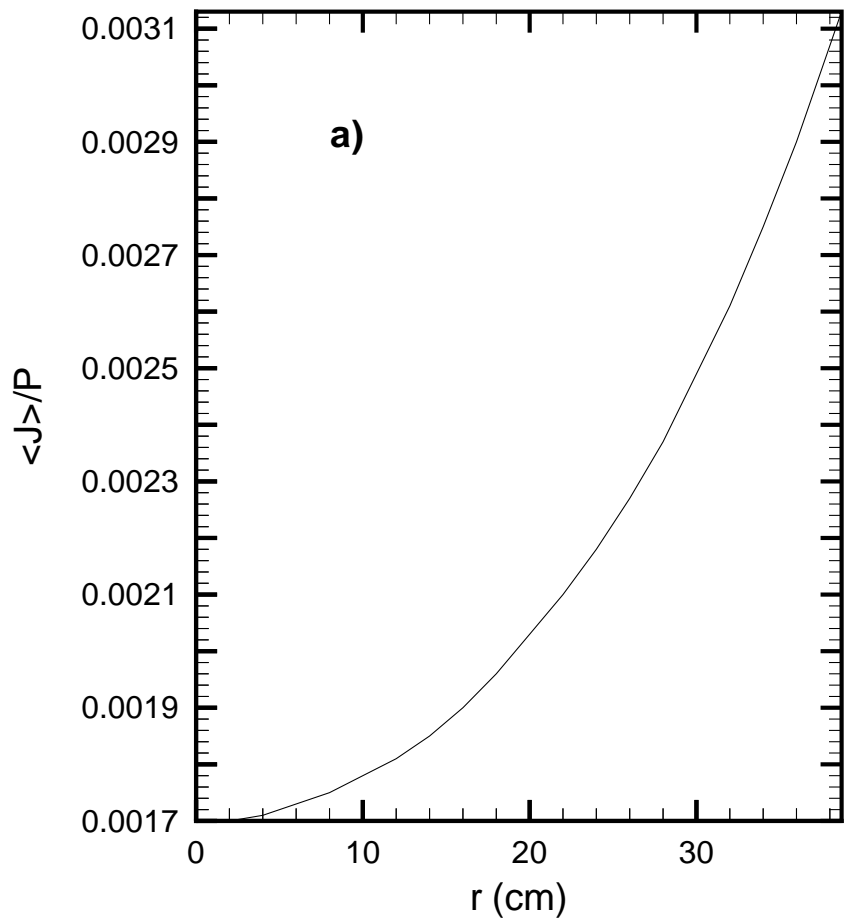
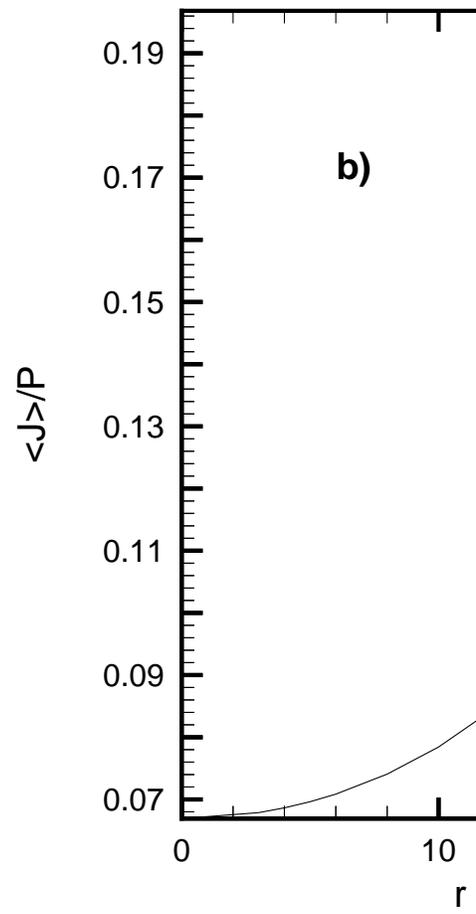
\end{document}